# Zero refractive index in space-time acoustic metamaterials

Theodoros T. Koutserimpas and Romain Fleury

*Laboratory of Wave Engineering, EPFL, 1015 Lausanne, Switzerland*

*Abstract:*

New scientific investigations of artificially structured materials and experiments have exhibit wave manipulation to the extreme. In particular, zero refractive index metamaterials have been on the front line of wave physics research for their unique wave manipulation properties and application potentials. Remarkably, in such exotic materials, time-harmonic fields have infinite wavelength and do not exhibit any spatial variations in their phase distribution. This unique feature can be achieved by forcing a Dirac cone to the center of the Brillouin zone ($\Gamma$ point), as previously predicted and experimentally demonstrated in time-invariant metamaterials by means of accidental degeneracy between three different modes. In this article, we propose a different approach that enables true conical dispersion at $\Gamma$ with twofold degeneracy, and generates zero index properties. We break time-reversal symmetry and exploit a space-time modulation scheme to demonstrate a time-Floquet acoustic metamaterial with zero refractive index. This behavior, predicted using stroboscopic analysis, is confirmed by fullwave finite elements simulations. Our results establish the relevance of space-time metamaterials as a novel reconfigurable platform for wave control.

## A. Introduction:

Near-zero refractive index metamaterials have been of great interest for researchers over the past few years [1–16]. In electromagnetics, such materials force Maxwell's equations to collapse to the Laplace differential equation (a special case of the Helmholtz equation) in the frequency domain, which effectively enforces a static wave profile to the time-harmonic electromagnetic field. Analogously, the wave equation for acoustic pressure in a zero-index fluid also transforms into the Laplace equation, and exhibits the same static wave behavior. It is therefore overt that waves in such artificial media experience infinite (or near infinite) phase velocity, hence they can be manipulated in very unique ways. For instance, taking into account the conservation of frequency (from the linearity of the wave equation) and the conservation of the parallel component of the wave vector $\mathbf{k}$ at an interface, such materials can bend, steer and focus the direction of a wavefront, ideally without any angular dependence, as illustrated in Fig. 1. In addition, they exhibit unique scattering and transmission properties when they are embedded in other kinds of media. For instance, in photonics, a non-magnetic ($\mu = 1$) zero-epsilon ($\varepsilon = 0$) photonic structure doped with a simple dielectric scatterer exhibits effective magnetic scattering features[7]. Both in acoustics and microwaves, a zero-index metamaterial (ZIM) can be engineered to function as a total reflector or a total transmitter depending on the parameters of the solid defect which is embedded within the artificial structure[8,11,17]. Such interesting properties were also recently introduced for electrons in artificial quantum lattices[16].

For such a material to exist, the dispersion relation must support a solution with near zero wavenumber at a non-zero frequency. One possibility to achieve this is to use periodic structures and create a Dirac cone at the $\Gamma$ point of the Brillouin zone[18–27]. Such phenomenon in electromagnetics has been predicted and experimentally verified by forcing the periodic structure to an accidental degeneracy point[19–21,23], obtained by tuning some geometrical parameters in the lattice. A recent article by M. Dubois, C. Shi et al[23] has used this technique to experimentally prove the existence of a Dirac cone at the center of the Brillouin zone of an acoustic metamaterial, and demonstrate a double zero refractive acoustic index with reasonably good impedance matching with air. In these systems, the Dirac cone is never a two-fold degeneracy, which would correspond to an effective spin of ½, but is instead a three-fold degeneracy (effective spin 1), and the conical dispersion is always accompanied by a third slow band with a different mode symmetry, consistent with the fact that two-fold degeneracies at the $\Gamma$ point must have parabolic dispersion in any time-reversal invariant crystal[28]. Nevertheless, these effective bosonic systems are very interesting as they lead to several exotic tunneling and statistical properties inherent to spin 1 particles[29,30]. In the present article, we follow a completely different approach to achieve near zero response, and demonstrate true linear (effective spin ½) dispersion at the center of the Brillouin zone by breaking time-reversal symmetry, turning a static lattice into a space-time crystal by applying a time-Floquet modulation.

### *B. The spatiotemporal periodic structure*

Initially introduced for quantum systems and studied in condensed matter physics[31–33], time-Floquet systems have been recently translated to the classical realm of wave physics. Time-Floquet metamaterials and space-time crystals have been used to induce novel wave phenomena, such as topological protection, nonreciprocity and space-time boundaries[34–41]. These structures correspond to systems whose index of refraction is not only modulated periodically in space, like for regular crystals, but also in time.

Let us consider the space-time metamaterial represented in Fig. 2. This Floquet acoustic crystal has already been discussed in another context for its unique and exotic topological properties[34]. It forms a hexagonal lattice constructed by coupled acoustic trimers connected together along the hexagonal bonds via small rectangular channels. The medium filling the acoustic crystal is silicone rubber RTV-602, which is a low loss material with density $\rho_0 = 990 kg \cdot m^{-2}$ and compressibility $\beta_0 = 9.824 \times 10^{-10} Pa^{-1}$ [42]. The surrounding medium is air, which effectively imposes hard wall acoustic boundary conditions to the rubber (large impedance mismatch), and the cavity diameter is 1 cm with a thickness of 3 mm. Following [43], at low frequencies the trimer can be viewed as a lumped element acoustic pressure network featuring three parallel coupled L-C resonators, with acoustic capacitance of $C_0 = \beta_0 V_0$, i.e. we operate the system way below the first dipolar resonant frequency of the individual cavities (circa 60kHz). We modulate periodically in time the acoustic capacitance of each cavity as show in Fig. 2a at a angular frequency $\omega_m = 2\pi f_m$, with a modulation depth of $\delta C$ and a phase $\varphi_m^i$ different for each symmetric cavity $i \in \{1,2,3\}$ as shown in Fig. 2a so that the modulation will convey an effective spin to the trimer and break time-reversal symmetry.

**The Static and Stroboscopic Analysis:** In the case of $\delta C = 0$ the system does not vary in time. The band structure of this case is shown in Fig. 3, where $k$ and $J$ correspond to the internal and external coupling coefficients of the structure (as indicated in Fig. 2a). From a standard tight binding model [44], it is straightforward to show that the distances of the bands at $\mathbf{k} = 0$ are directly dependent on the relative values of the internal and external coupling coefficients, as shown in the figure. The next step of the approach is to consider the effect of nonzero time-

Floquet modulation $(\delta C \neq 0)$. In order to do this, we calculate the stroboscopic (Floquet-effective) Hamiltonian of the system:

$$H_{eff} = \frac{j}{T} \log U(T), \qquad (1)$$

where:

$$U(t) = \mathsf{T} \, \exp\left[-j \int_0^t H(t)dt\right], \qquad (2)$$

and $\mathsf{T}$ is the time ordering operator. This effective Hamiltonian has the ability to describe the time evolution of the system at discrete times separated by $T$, in a similar way as a stroboscopic photography can sample the motion of a moving item. The dynamics of our system are actually much simpler than the ones of an arbitrary stroboscopic photographic shot: indeed, our Hamiltonian is not only time-dependent, but also time periodic with a time period of $T_m = 2\pi/\omega_m$. This information is crucial, because it implies that we can define a quasi band-structure that repeats itself along the frequency axis (Floquet theorem[45]) every $\omega + n\omega_m$ (where $n$ is an integer). The repartition of the modal energy among the various Floquet Harmonics describes the frequency content of the Floquet mode. Next, we show the effect of the modulation on the quasi-band structure of the system.

To simplify the design procedure, we start by isolating the upper two bands of the quasi band structure from the bottom four, in the limit of vanishingly small modulation depth. We will then focus only on what happens to the bottom four bands when the modulation depth is increased. For this purpose, let us define the dimensionless parameters: $Q_T = \omega_0/2J$, $Q_M = \omega_0/2k$, $x_T = Q_T \omega_m/\omega_0$ and $x_M = Q_M \omega_m/\omega_0$ where $\omega_0$ is the resonant frequency of the cavities and

$J, k, \omega_m$ the external coupling, the internal coupling and the modulation frequency, respectively. From the tight binding model[34], it is straightforward to show that the condition that guarantees that the two upper bands remain isolated from the rest once folded along the frequency axis is:

$$\frac{3}{2}\frac{x_T}{x_M} + 1 - \left[1 + \left(\frac{3}{2}\frac{1}{x_M} - \frac{1}{x_T}\right)\right] x_T < 0. \tag{3}$$

We also stress that the physical meaning of the parameter $x_T$ is to quantify the empty frequency space left (not occupied) in the temporal Brillouin zone by the bands. Values of $x_T$ larger than unity mean that some space is left in the temporal Brillouin zone, and under this condition only it is possible to find values of $x_M$ that will not fold the two upper bands on top of the four lower bands. In order to demonstrate the sway that the time-Floquet dynamics dictate to the system we set our structure with the following parameters (as an example): $\omega_0 = 200$MHz, $\omega_m = 17.89$MHz, $J = 2$MHz, and $k = 62.6$MHz which result to $Q_T = 50$, $Q_M = 1.6$, $x_T = 4.5$, and $x_M = 0.1428$ (these values satisfy the condition set at Equation (3)) and we also define a dimensionless parameter that represents the depth of the time-modulation: $y = Q_T \delta\omega/\omega_0$ (where $\delta C/C_0 = \delta\omega/\omega_0$). It is evident that when we do not apply any time modulation ($y = 0$) the stroboscopic analysis coincides with the static analysis (as shown in Fig. 4a, which illustrates the same band structure as in Fig. 3, although folded along the frequency axis). As we increase the modulation depth, the upper two bands are not affected by the activation of the Floquet dynamics, due to the monopolar symmetry of the modes, which does not overlap with the dipolar modulation scheme. However, the rest of the bands (see green arrows) are strongly affected by the modulation, and become very flat, opening large bandgaps (as shown in Fig.4b for $y \approx 2.12$), a phenomenon which has been utilized to create topological sound insulators[34]. Yet, if we continue to tune up the level of the modulation (for $y \approx 3.134$), another interesting phenomenon occurs. The two bands highlighted with light green arrows in Fig. 4 get close to each other at the center of the Brillouin zone and form a Dirac cone, turning the sound insulator into a time-space zero index acoustic metamaterial at the Dirac frequency. Fig. 4c shows the 3D graphic representation of the two bands of interest, demonstrating the existence of genuine conical dispersion over almost the entire Brillouin zone. It is remarkable that the dispersion of the massless Dirac phonons remains linear over a very large area of the Brillouin zone.

### C. The fullwave simulation

So far, our investigations were only based on a toy model relying on the tight-binding stroboscopic Hamiltonian. We now turn to fullwave finite element simulations to corroborate the exact behavior of the system as a stroboscopic zero index structure. We set the parameters $\delta C / C_0 = \delta \omega / \omega_0 = 9.745\%$, with a time-modulation frequency of $f_m = 1312.5$Hz, and coupling rates $k \approx 6840$Hz, and $J \approx 209.38$Hz. It is verified that equation (3) stands also for the simulation. The computational tool we use to illustrate the time-Floquet dynamics of the system is the finite-element method with a truncation of the Floquet expansion of the acoustic wave equation to -1, 0 and +1 Floquet Harmonics[43].

At first, we compute the dispersion relation of the structure. Our results shown in Fig. 5a show the quasi-band structure obtained in the simulation for the four bands of interest and the two upper bands are indeed isolated (not shown in Fig. 5 for brevity). The graph of the dispersion follows the theoretical results of the band structure at Section B. Remarkably, the results are in excellent agreement with the analytical prediction, demonstrating the expected conical dispersion for frequency $f = 21889$Hz at $\mathbf{k} = 0$. Fig. 6b shows the three dominant Floquet Harmonics associated with the two degenerate acoustic modes at $\Gamma$. The dominant frequency at which we have wave propagation in this system is the 0 Floquet Harmonic. This means that there is no significant energy conversion to adjacent Floquet Harmonics, consistent with the relatively low value of the modulation depth.

Since we know the band structure of our model, we can now simulate acoustic pressure propagation in finite-size acoustic metamaterials, exciting the stroboscopic zero refractive index behavior for the specific Dirac frequency $f = 21889$Hz (Fig. 7 and 8). To excite finite sized systems with plane waves, we used two rectangular domains (labeled I and II in Fig. 6, 7), filled with water and located outside the metamaterials. Both the simulations in Fig. 6, 7 depict the scattering profile of the acoustic metamaterial for -1, 0 and +1 Floquet Harmonic acoustic pressure fields. More specifically, in Fig. 6 the metamaterial forms a slab: the incident and reflected fields are illustrated in domain I and the transmitted field is illustrated in the domain II. Fig. 7 shows a similar simulation for which the metamaterial has the shape of a prism. For the simulation of Fig. 7 it is shown that the transmitted field is actually a steered plane wave with transmitted wavefront determined by the direction of the output interface, a result which is in full agreement with the propagation theory of zero-index materials as described in Fig. 1a. Looking at the scale bar, it is clear that the dominant frequency of the system remains at the 0 Floquet Harmonic, which corresponds to the frequency of the incident acoustic field.

**Characterization of the material:** We now move to a direct quantitative demonstration of zero index properties using the fullwave simulations. The simulation of a finite-thickness slab depicted in Fig. 6 can be used to extract an effective refractive index ($N_{eff}$) for the metamaterial and an effective acoustic impedance ($Z_{eff}$). The computation of the model of Fig. 6 gives us the transmission ($S_{21}$) and reflection ($S_{11}$) of the field. With these simulation results, it is straightforward to determine the characteristics of the material as a homogenous medium, by applying the equations [25,46,47]:

$$Z_r = \frac{Z_{eff}}{Z_0} = \pm\sqrt{\frac{(1+S_{11})^2 - S_{21}^2}{(1-S_{11})^2 - S_{21}^2}}, \quad (4)$$

$$\exp(jN_{eff}k_0 d) = \frac{S_{21}}{1 - r\frac{Z_r - 1}{Z_r + 1}}, \quad (5)$$

where $d$ is the slab thickness. These formulae give an acoustic impedance of $Z_{eff} = (295.86 + j1050.7)k\Omega$ and an effective acoustic index of $N_{eff} = -0.0127 + j0.0056$. As expected, the real part of the relative acoustic index is very close to zero. The non-zero imaginary part represents losses from the viewpoint of the zeroth order Floquet field: these losses are caused by the energy leakage to its neighbor Floquet harmonics, which carries away a small albeit noticeable amount of energy.

### *D. Conclusion*

In this work, we have presented an active acoustic metamaterial that exhibits near zero refractive index. In contrast with the existing literature, we derived this peculiarity by activating time-Floquet dynamics and by tuning the time-modulation depth to a specific point that provides a quasi-band structure featuring a Dirac cone at the $\Gamma$ point. Such structure possesses the advantage to be reconfigurable and to provide a unique platform for observation of acoustic waves with pseudo-spin ½. We believe that this study enriches the pre-existing research, highlighting the fascinating effects that a space-time modulation can impart to a given structure and the extraordinary properties of waves in time-modulated acoustic, electromagnetic or mechanical metamaterials. This work was supported by the Swiss National Science Foundation (SNSF) under Grant No. 172487.

# Figures

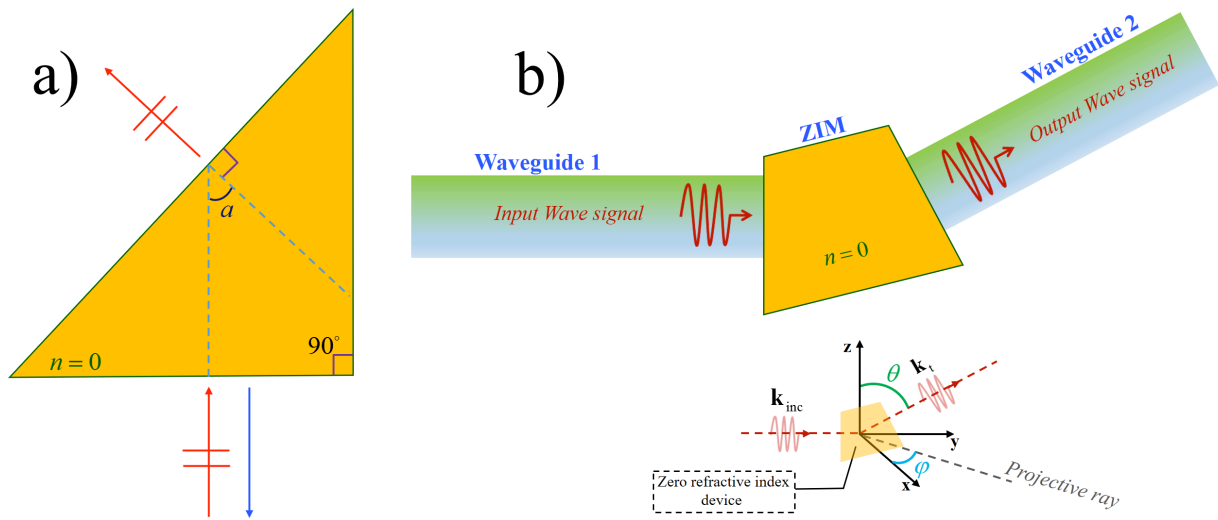

**Fig. 1: Graphical illustrations of the exotic properties of zero index materials.** a) The propagation of a ray in a zero-index prism, b) a zero-index device which redirects a wave signal in between two waveguides.

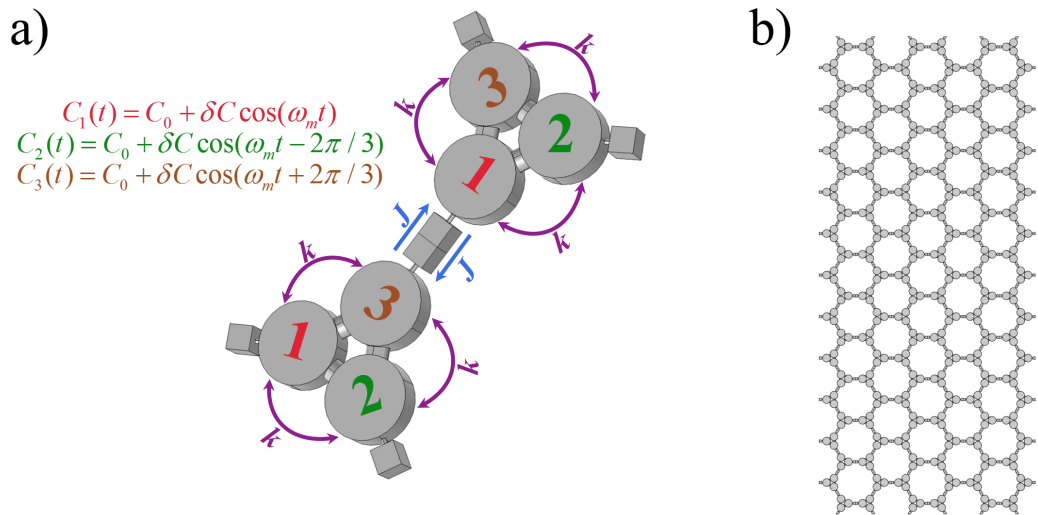

**Fig. 2: The considered spatiotemporal periodic structure.** a) the unit cell of the periodic structure with the time modulation of the acoustic capacitance and the internal and external coupling coefficients, b) the complete periodic structure from a macroscopic point of view. The grey domains are filled with silicone rubber. The external medium is air.

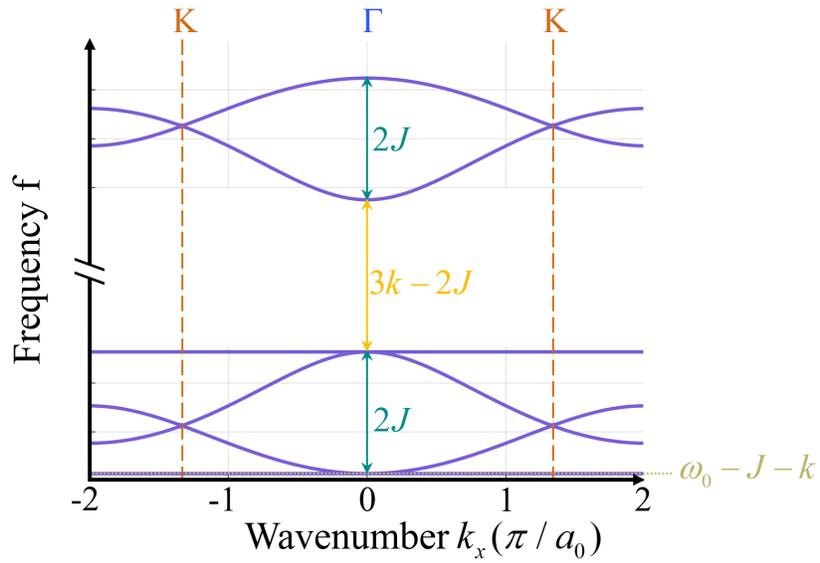

**Fig. 3: The band structure of the static crystal.** The top four bands are of dipolar nature, whereas the top two bands are of monopolar nature.

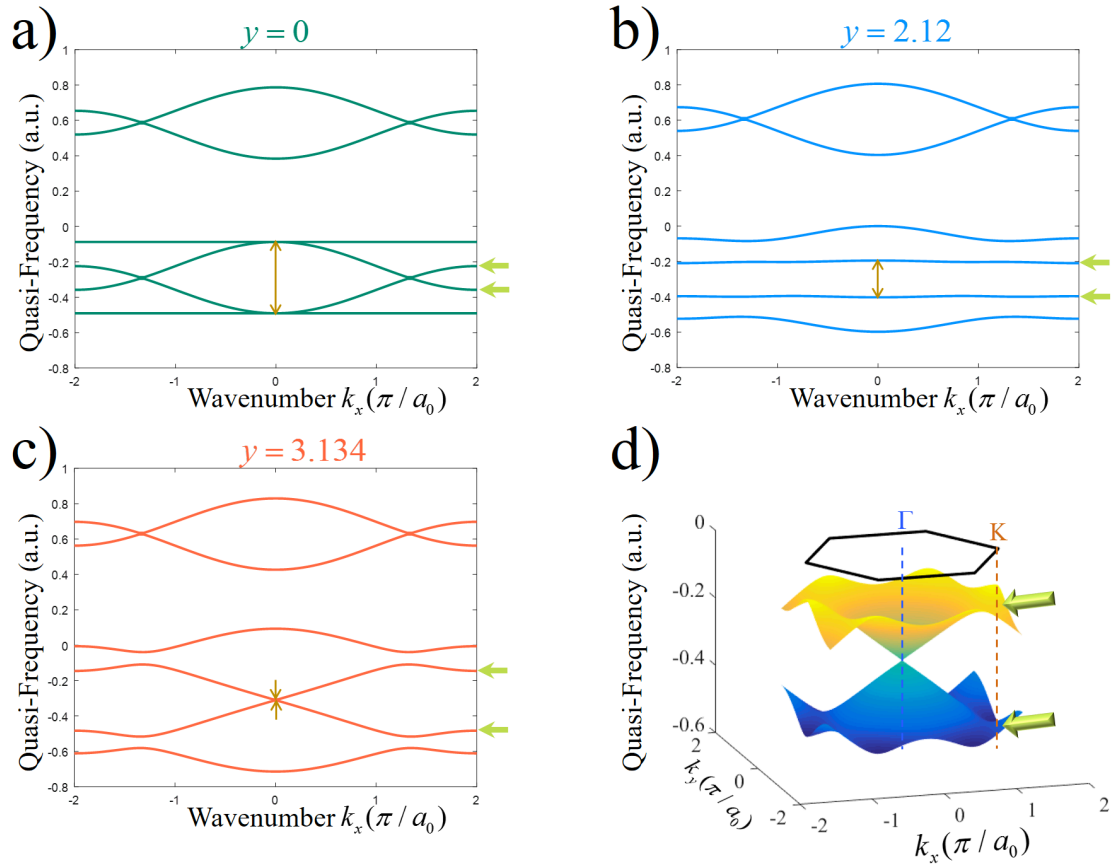

**Fig. 4: The quasi band structure of the time-Floquet system as the modulation depth increases** (the two bands that eventually form a Dirac cone at $\Gamma$ are highlighted by the light green arrows). a) the band structure of the unmodulated system ($y = 0$) (same as in figure 3, but folded), b) the band structure becomes flat as modulation increases ($y \approx 2.12$), c) at a specific modulation depth the Dirac cone appears in the center of the Brillouin zone ($y \approx 3.134$), d) a 3D graphical representation of the Dirac cone, showing its large spatial extent in the Brillouin zone (black hexagon).

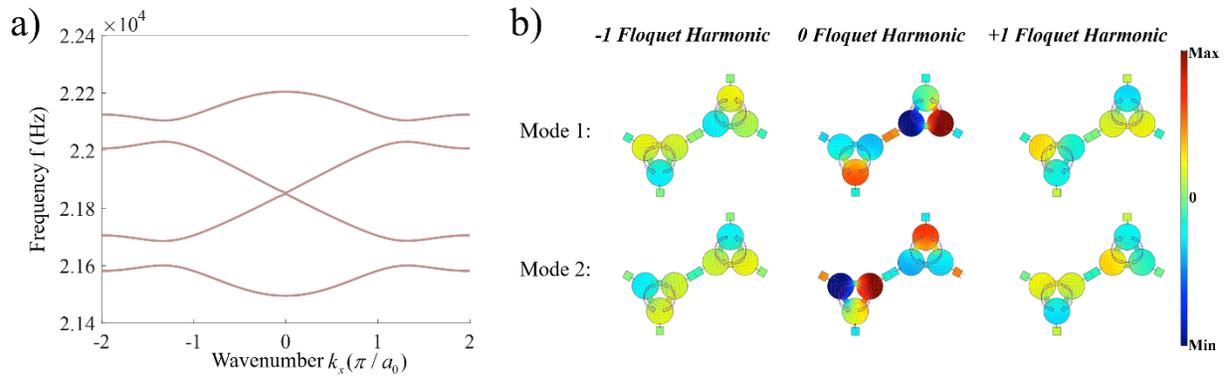

**Fig. 5: FEM simulation of the infinite space-time crystal.** a) Band structure of the fullwave simulation for $\delta C/C_0 = \delta\omega/\omega_0 = 9.745\%$, and $f_m = 1312.5\text{Hz}$, and the formulation of the Dirac cone at $f = 21889\text{Hz}$, b) acoustic pressure of the modes at the Dirac frequency $f = 21889\text{Hz}$ (frequency components corresponding to the -1, 0 and +1 Floquet Harmonics).

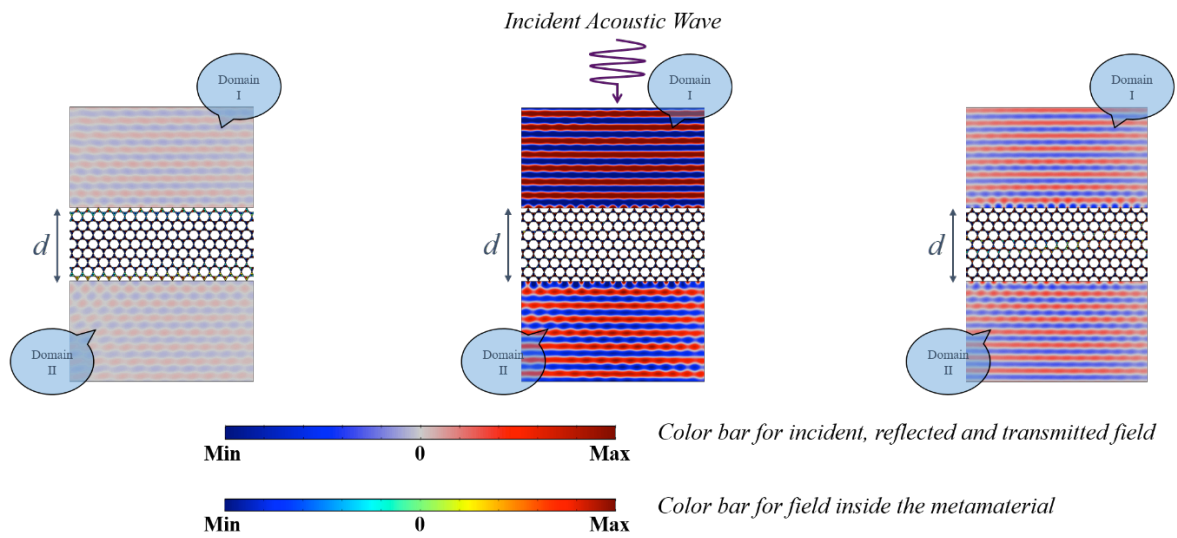

**Fig. 6: Fullwave simulation of a zero-index time-Floquet slab.** Acoustic pressure field as it is reflected and transmitted by a rectangular metamaterial slab (domains I, II are filled with water).

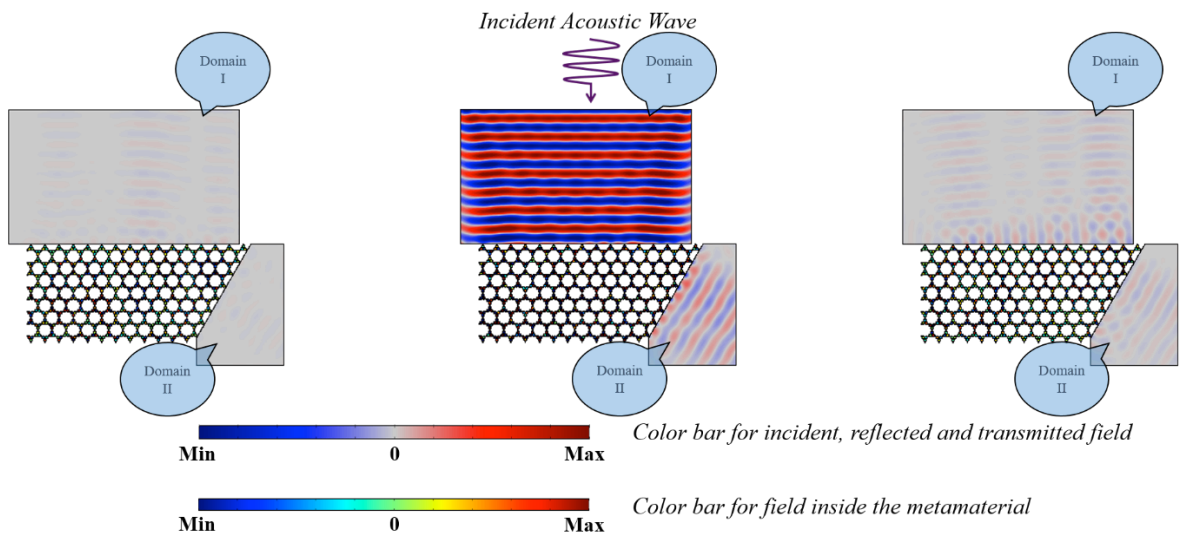

**Fig. 7: Fullwave simulation of a zero-index time-Floquet prism.** Acoustic pressure field as it is reflected and transmitted by a metamaterial prism, demonstrating zero-index properties at the fundamental harmonic frequency (domains I, II are filled with water).